\documentstyle[12pt]{article} \hoffset -0.5in \textwidth 6.0in \textheight
8.00in  \parskip 7pt \openup4.0\jot \parindent=0.5in
\input{[mcb]macros.tex}
\renewcommand{\thefootnote}{\fnsymbol{footnote}}

\def\mainhead#1{\setcounter{equation}{0}\addtocounter{section}{1}
  \vbox{\begin{center}\large\bf #1\end{center}}\nobreak\par}

\supercite

\begin{document}
\textheight 9.0in
\textwidth 6.5in
\begin{titlepage}
\rightline{\vbox{\halign{&#\hfil\cr
&ANL-HEP-PR-92-25\cr
&February 1992\cr}}}
\vspace{1in}
\begin{center}

{\Large\bf
Phenomenology of R-parity Breaking in $E_6$ Models}\footnote{Work supported
by the U.S. Department of
Energy, Division of High Energy Physics, Contracts\newline
W-31-109-ENG-38 and W-7405-Eng-82.}
\medskip

\normalsize Thomas G. Rizzo
\\ \smallskip
High Energy Physics Division\\Argonne National
Laboratory\\Argonne, IL 60439\\
and\\
Ames Laboratory and Department of Physics\\
Iowa State University, Ames, IA 50011\\
\end{center}

\begin{abstract}
We explore the phenomenology of new R-parity violating operators that can occur
in $E_6$ models.  The set of allowed operators is found to depend   sensitively
on the nature of the extension of the standard model gauge group.  These new
interactions lead to additional production processes for the exotic particles
in such models and allow the LSP to decay but with a highly suppressed rate.
The implications of these new interations are
examined for the Tevatron, SSC, LHC, HERA, and
$\sqrt{s} = 0.5$ and $1$ TeV $e^+e^-$ colliders.
\vspace{0.75in}
\end{abstract}

\medskip
\renewcommand{\thefootnote}{\arabic{footnote}} \end{titlepage}
\textheight 8.0in
\textwidth 6.0in

It is by now well known that in the minimal SUSY standard model gauge
symmetries alone do not forbid the existence of baryon ($B$) and/or lepton
$(L)$
number violating term in the superpotential which break R-parity and lead to a
rich phenomenology.$^1$  Of course, if rapid proton decay is to be avoided, not
all of the terms allowed by gauge invariance can occur simultaneously so that
additional symmetries must be present forbidding either the  $\Delta B \neq 0\
\rm {or}\ \Delta L \neq 0$ terms.

In $E_6$ models, the enlargement of the particle spectrum from the usual
15 to 27
2-component fields per generation already leads to a fairly complicated
structure for the superpotential$^2$ when R-parity is conserved. In
fact, even in this case not all of the usual 11 terms in the superpotential
can be
allowed simultaneously if we wish to ensure proton stability.  Another way to
think about this is that one finds that the $B$ and $L$ assignments for all
of the
`exotic' fields in the \underline {27}\ representation of $E_6$ will not be
uniquely defined if all 11 terms of the superpotential are present.  For
example, we choose whether the color-triplet, isoinglet, $Q = -1/3$ field is a
leptoquark, diquark, or an ordinary quark by which terms we  allow in the
superpotential.  If we are willing to give up R-parity in these extended
 models this
kind of complexity will be only significantly compounded.  Of course in
general $E_6$ models, the low energy gauge symmetry is larger than that of
the SM by at least an additional U(1) factor. This additional symmetry may
disallow several if not
all of the new terms in the superpotential which violate R-parity.  The
structure of the new terms, whether they are allowed in models with particular
extended
gauge symmetries, and some of the possible phenomenological implications will
be
the main points we wish to address in this paper. We will restrict our
attention to the existence of explicit, renormalizable R-parity violating
terms in the superpotential in our discussion below. The possibility of
introducing non-renormalizable higher dimension terms induced by loops will be
discussed elsewhere.

We know in the MSSM that requiring only $SU(3)_c \times SU(2)_L \times U(1)_Y$
gauge invariance and renormalizability leads not only to superpotential ($W$)
terms which generate
fermion masses and Higgs self-couplings
$$
W_1 = Qu^c\ H^c + Qd^c\ H + Le^c\ H + HH^c
\auto
$$
but also to $\Delta L,\ \Delta B \neq 0$ terms as well:
$$
W_2 = QLd^c + LLe^c + u^cd^cd^c + LH^c
\auto
$$
where coupling constants and generation indices   are suppressed.  The first,
second, and fourth terms in $W_2$ violate $L$ whereas the third term violates
$B$;
the $LH^c$ terms can in principle be eliminated by a suitable rotation among
the
superfields.$^1$  As mentioned above,  {\it both}\ $\Delta L \neq 0$ and
$\Delta B \neq 0$ terms cannot occur simultaneously and this is usually
insured by the existence of
a new discrete symmetry.

In $E_6$ models imposing  {\it only}\ $SU(3)_C \times SU(2)_L \times
U(1)_Y$ gauge symmetry  leads to a large increase in the number of terms in
$W$.
In order to fix the notation for the exotic fields in the \underline {27}\
representation we follow the nomenclature used in Hewett and Rizzo$^2$ and
shown in Table~1.  One finds that there are now 25 possible trilinear terms
(shown
in Table~2) and 7 possible bilinear terms (shown in Table~3).  Of these 32
terms, 11 are those usually associated with $W$ in $E_6$ models, 4 are the
R-parity violating terms in (2) and one is the $HH^c$ term in (1); thus 16 new
terms are generated which contain `exotic' fields, \ie, those present in the
\underline {27}\ of $E_6$ but not in the MSSM.  We note that all of the new
trilinear terms contain {\it at least two}
exotic superfields.  In what follows, we
will generally ignore any new physics associated with the terms violating
R-parity in the MSSM which appear in (2) since they have already been
examined in the literature$^1$ and will already be familiar to the reader.

Before discussing the phenomenology of the new terms in $W$ we remind the
reader
of the influence of the terms which already appear in the standard R-parity
conserving version of $E_6$ models.  For example, terms 4, 5, 8, 23, and 24 in
Table~2 are responsible for generating the masses of $u-$ and $d-$type quarks,
the charged leptons and the exotic fermions $H$ and $h$,
respectively.  If any of the terms
3, 13, and 15 in Table~2 are present, then $B(h) = 1/3\ \rm {and}\ $L$(h= -1)$,
\ie, $h$ is a leptoquark; whereas if terms 1 or 12 in Table~2 are present,
then $B(h) = -2/3$\ and $L(h) = 0$, \ie, $h$ is a diquark. (If none of the
terms are present, we can identify $h$ as an ordinary quark.)  As is well
known, these two sets of terms in $W$ cannot exist simultaneously since then
$h$ would have ill-defined $B$ and $L$ quantum numbers and can thus mediate
proton
decay at a substantial rate unless $h$ has a mass of order the unification
scale $\approx 10^{14}$ GeV or greater.  We will consider the effect of the
new terms in
$W$ within the context of these two possible quantum number assignments.
Lastly, we note that $B(H) = L(H) = 0 = B(S^c) = L(S^c)$ since the
corresponding neutral
scalar field components act as Higgs
and obtain vacuum expectation values (vevs); in addition
 $B(\nu^c) = 0$ whereas
$L(\nu^c) = 0\ \rm {or}\ -1$ depending on the model, \ie, whether
$\tilde{\nu}^c$ must have a large vev in order to break any additional gauge
symmetries down to the SM.

Which of these terms in $W$ would survive if extended gauge symmetries are
present in the limit that such symmetries are unbroken?
The answer, of course, depends on the nature of the additional
symmetry.  As an example of an additional U(1) symmetry, consider the breaking
$$
E_6 \rightarrow SO(10) \times U(1)_\psi \rightarrow SU(5) \times U(1)_\chi
\times U(1)_\psi
\rightarrow {\rm SM}  \times U(1)_\theta
\auto
$$
where $U(1)_\theta \equiv U(1)_\psi \cos \theta - U(1)_\chi \sin \theta$ is the
additional gauge symmetry with $-90^\circ \leq \theta \leq 90^\circ$.  Clearly,
as $\theta$ is varied and the couplings of the superfields change different
terms in $W$  will be allowed.  Of course, the 11 trilinear terms in $W$
present in R-parity conserving models will occur for all $\theta$ values.
Tables~2 and 3 show which terms are allowed for various values of the
parameter $\theta$.
For example,
model $\eta (\theta = 37.76^\circ)$ allows the most number of new terms
including the all of the usual ones of the MSSM with R-parity breaking,
whereas model $\chi (\theta = -90^\circ)
$
only allows the $(S^c)^3$ term among the trilinear couplings and terms 5--7
among the bilinears.  For {\it arbitrary}\ values of $\theta$,
{\it no}\ R-parity violating terms are found to be allowed by the
additional gauge symmetry and, thus, all bilinear
couplings vanish.  Thus, in general, the additional $U(1)_\theta$ symmetry
removes the possibility of explicit
R-parity violation {\it except}\ for specific
values of the $\theta$ parameter.

What about other gauge extensions?  Let us consider the extension to the
$SU(3)_c \times SU(2)_L \times SU(2)_R \times U(1)$ model; there are two
versions of this scenario within the $E_6$ context.  The first, which is the
conventional left-right model (LRM)$^3$ places the usual right-handed quarks
and leptons in right-handed doublets and, \eg, $h_L, h_L^c$ are both $SU(2)_L
\times SU(2)_R$ singlets.  There is, however, a second possible quantum
number assignment due to Ma and collaborators$^4$ which is an alternative
version of the LRM (ALRM).  Basically, the quantum number assignments of the
pairs  $L \leftrightarrow H, d^c \leftrightarrow h^c$, and  $\nu^c
\leftrightarrow S^c$ are interchanged in this scheme.  Imposing this extended
gauge symmetry, the only new trilinear term to survive is  $(S^c)^3
[(\nu^c)^3]$
for the ${\rm LRM} [{\rm ALRM}]$ case while three bilinears are found to
be allowed for each of the two
scenarios.

{}From this exercise we conclude that the existence of additional gauge
symmetries greatly reduces the chance that explicit
R-parity violating terms will be
present in the superpotential of $E_6$ models.  We now turn to a brief
discussion of the phenomenology of the various new R-parity violating terms
that can appear in $W$.  Hattori \etal $^5$ have already addressed the issues
of
chargino masses and generating neutrino mass hierarchies in $E_6$ models with
R-parity violation so we will not deal with those subjects here, although
further work in both areas clearly still needs to be done.  Thus we will, for
now,
avoid operators which are most important for dealing with the problems of
neutrino mass generation.

We note, of course,that although additional gauge symmetries may forbid the
existence of renormalizable tree-level operators in the superpotential which
break R-parity,such terms can be generated at the one- or two-loop level thus
resulting in rather small effective Yukawa couplings. In the LRM,for example,
the existence of a vev for the scalar partner of $\nu^c$ implies that one can
write down effective dimension-5 operators which are gauge invariant and
R-parity conserving but lead to the conventional R-parity violating terms in
the MSSM due to a spontaneous breaking of R-parity. This point has been
stressed by Valle and collaborators $^1$.

The first trilinear R-parity violating operator not occurring in the MSSM  is
$QHh^c$.  Since $L(H) = B(H) = 0$ this operator is $|\Delta B| = 1$ if $h$ is a
diquark whereas, if $h$ is a leptoquark, this operator is $|\Delta B| = |\Delta
L| = 1$ which could allow for rapid proton decay in conjunction with the usual
R-parity conserving operators in $W$.  (We note that this operator, like all of
the new R-parity violating trilinears, contains two exotic superfields.)  With
$H^T = (N,E)$, the operator decomposes into $dNh^c + uEh^c$, where
$Q(N,E) = (0, -1)$, following the notation of ref. 2.  This operator
would, \eg, allow for the associated production of exotic particles at hadron
colliders which were not particle-antiparticle pairs.  Conventionally,$^2\ h$'s
are pair produced via $qq, gg \rightarrow h\overline{h}(\tilde{h}\bar{\tilde
h})$  but we
can now have both $gu \rightarrow \tilde{E}h(E\tilde{h})$ and
$gd \rightarrow \tilde{N}h(N\tilde{h})$ reactions as well.  Since $H$'s are
color singlets, they can only be produced via purely electroweak interactions
in the R-parity conserving case but the existence of this new operator,
 $QHh^c$,
could allow for their production with much larger rates at hadron
supercolliders as we will see below.
Note that the operator $d^cS^ch$ leads to new physics similar to $QHh^c$
(except for the replacement $N \rightarrow S$) since it can lead to  $gd
\rightarrow \tilde{S}h(S\tilde{h})$ but there is no comparable $gu$ initiated
process.

The operator $LH^cS^c$ is $|\Delta L | = 1$ since $B (H,S) = L(H,S) = 0$ and
decomposes into $\nu N^cS^c + eE^cS^c$.  As in the previous case, this operator
allows for the associated production of exotics, not at hadron supercolliders,
but at $\epm$\ colliders via the process $\gamma e \rightarrow
E\tilde{S}(\tilde{E}S)$.  The rates for this process can also be quite large as
we will see below.  The operator $e^cHH$ leads to similar physics since it
generates the interaction $e^cNE$ but without a comparable term involving all
neutral fields.

The last trilinear operator is $u^ch^ch^c$.  If $h$ is a diquark, then this
term is $|\Delta B| = 1$ whereas if $h$ is a leptoquark, both $\Delta B$ and
$\Delta L \neq 0$ interaction are induced.  This type of operator can lead to
$h\tilde{h}$ production in hadronic collisions via the process
$gu \rightarrow h\tilde{h}$.  We remind the reader that the conventional $gg$
and $q\overline{q}$ production   mechanics lead to $h\overline{h}$ or
$\tilde{h}\bar{\tilde h}$ and not to a mixed $h\tilde{h}$ final state.

Among the bilinear operators almost all of them simply induce bare mass terms
for $\nu^c,S^c,H$ and $h$.  The $hd^c$ term parallels the usual $LH^c$ of the
MSSM which induces mixing between the two superfields; in the $hd^c$ case,
identifying $h$ as a diquark (leptoquark) this operator is $|\Delta B| = 1$
($\Delta B, \Delta L \neq 0$).  As in the $LH^c$ case, too, this term can be
rotated away by a suitable redefinition of the fields.  (Note that such a
rotation may not always possible if additional gauge symmetries are present.)
Thus, except for mass generation, there is not much additional physics
associated with the new bilinear terms.

One of the `hallmarks' of R-parity violating models is the instability of the
$LSP$, the lightest SUSY partner.  In the MSSM (assuming the $LSP$ is the
lightest neutralino, $\chi^0_1$)  this particle
can undergo a 3-body decay, via, \eg, the
$LLe^c$ term in Eq.~(2), \ie, $\chi^0_1 \rightarrow \overline{e}\tilde{e}^
*\rightarrow \overline{e} e \nu$.  Depending on the size of the $LLe^c$
coupling, $\chi^0_1$ may decay inside the detector$^1$ and invalidate the usual
SUSY signal of missing energy.  If we do {\it not}\ consider the MSSM
R-parity violating terms in $W$ here, we find that the $\chi^0_1$ decay will be
very highly suppressed in comparison to the MSSM case since it must proceed
through virtual exotics.  For example $\chi^0_1 \rightarrow
h^* {\bar{\tilde h}}^*$,
followed by ${\bar{\tilde h}}^* \rightarrow q\overline{q}$, $h^* \rightarrow
(N,S)^*d$, and $(N,S)^* \rightarrow q \overline{q}$.  Thus
instead of an effective dimension-6 operator responsible for $\chi^0_1$ decay
in
the MSSM case with R-parity violation
we find an effective dimension-9 operator for $E_6$ models.
So, in the MSSM case one
obtains $\tau_\chi \sim m^5_\chi$ whereas in $E_6$
models $\tau_\chi \sim m^{11}_\chi$!  In fact, scaling unknown
Yukawa couplings to the
electromagnetic strength and taking the effective scale of the dim-9 operator
to be $\Lambda$ we estimate
$$
\Gamma_\chi = \lambda\ {{\alpha^4} \over {2048\ \pi^3}}\ m_\chi \
\left( {{m_\chi} \over {\Lambda}}\right)^{10}
\auto
$$
where $\lambda$ accounts for our ignorance of the Yukawas and the deviation of
5-body phase space integration of the matrix element from uniformity.  The
$\chi^0_1$ decay length is then given by
$$
d = 4.42.10^{-3} cm\ \gamma\beta\lambda^{-1}
\left({{m_\chi} \over {100\ {\rm GeV}}}\right)^{11}
\left({{\Lambda} \over {100\ {\rm GeV}}}\right)^{10}
\auto
$$
Fig.~1 shows $d$ as a function of $m_\chi$ assuming $\Lambda = 100$ GeV for
$\chi$'s produced
 with an energy of $100$ GeV and different choices of $\lambda$.
We see that if $m_\chi \leq 40$ GeV, even
with $\lambda = 1$, the LSP will escape
the detector before decaying.  For larger masses,
whether $\chi^0_1$ will have an
observable decay depends critically on $\lambda, \Lambda,\ {\rm and}\ m_\chi$.

Let us now turn to the various production processes for exotics discussed above
that are now possible given an R-parity violating $W$ arising from
$E_6$ models.
We first consider the case $\gamma e^\pm \rightarrow E^\pm \tilde{S}$ at
a high-energy $\epm$\ collider; the corresponding $S\tilde{E}^\pm$  cross
section
is then obtainable by interchanging the roles of the two masses, $m_E$ and
$m_S$, in the expressions below.  (Similarly $\gamma e^\pm \rightarrow E^\pm
\tilde{N}$ can be trivially obtained.)
Of course, the real process we are studying
is $e^+e^- \rightarrow E^\pm \tilde{S} e^\pm$ where the $e^\pm$ is unobserved
and the initial state photon comes from one of the original $e^\pm$.

The subprocess differential cross section for this reaction can be obtained
with
$(\hat{s} \equiv xs)$ where $x$ being the fraction of the initial $e^\pm$
energy
carried by the photon:
$$
{{d\hat{\sigma}} \over {dz}} = {{\pi \alpha^2} \over {\hat{s}^2}}\ \kappa
\beta\
\left[ A_0\, \tilde{u}^{-2} + A_1 \tilde{u}^{-1} + A_2 - \tilde{u}\right]
\auto
$$
where $\kappa \equiv (\lambda/e)^2,\,
z = \cos \theta,\, \tilde{u} \equiv \hat{u}-m^2_E =
- \sqrt{\hat{s}} \, (E_E-p_Ez)$, and
$$
\eqalign{
A_0 &= -2\hat{s} m^2_E\, (m_E^2 - m^2_S)\cr
A_1 &= - \left[ \hat{s}^2 + 2\hat{s} (m^2_E - m^2_S) + 2(m^2_E - m^2_S)^2
\right]\cr
A_2 &= -2(\hat{s} + m^2_E - m^2_S)\cr
\beta &= \left[ \left(1- {{m^2_E + m^2_S} \over {\hat{s}}} \right)^2 -
{{4m^2_Em^2_S} \over {\hat{s}^2}}\right]^{{1} \over {2}}\cr
}
\auto
$$
Here, $\lambda$ is the {\it a priori\/} unknown $e^cE^cS^c$ coupling.  Eq.~6
can be integrated to yield the sub-process total cross section.
$$
\hat{\sigma} = {{2\pi\alpha^2} \over {\hat{s}^3}}\ \kappa \
\left[ A_0(u^{-1}_l-u^{-1}_u)+ A_1 \ell n {{u_u} \over {u_l}} + A_2
(u_u-u_l)- {{1} \over {2}} (u^2_u-u^2_l)\right]
\auto
$$
where $u_{u,l} \equiv - \sqrt{\hat{s}} (E_E \pm p_E z_0)$ with $-z_0 \leq z
\leq z_0;
P_E = \sqrt {E^2_E-m^2_E}$\ and
$$
E_E = {{\hat{s} + m^2_E-m^2_S} \over {2\sqrt{\hat{s}}}}
\auto
$$
(For numerical purposes we take $z_0 =1$ in our calculations)
The total cross section is then, summing over both $e^+$ and $e^-$ initial
states,
$$
\sigma = 2\int^1_{x_L}\ dx\ f_{\gamma/e}(x)\ \hat{\sigma}(x)
\auto
$$
with $f_{\gamma/e}(x)$ being the photon flux and $x_L \equiv (m_E+m_S)^2/s$.
In what follows we will calculate $\sigma$ for $\epm$\ colliders with
$\sqrt{s} = 0.5$ or $1$ TeV and using either the photon flux of the
Weizsacker-Williams Effective Photon Approximation (EPA)$^6$
 or the beamsstrahlung (BS) photon$^7$ flux.  Figs.~2a--d show $\sigma$ as a
function of $m_E$ for different choices of $m_S$ assuming $\kappa
 = 1$.  Note that
for smaller values of $m_{E,S}$ the BS spectrum produces a larger value of
$\sigma$ in comparison to EPA whereas the reverse is true for large masses.  As
an example of the rates one should expect consider a $\sqrt{s} = 500$ GeV
(1 TeV) collider with an integrated luminosity of $25 (100) fb^{-1}$ and
take $M_E=M_S= 200(400)$ GeV.  For the EPA spectrum we obtain 42 (141) events
whereas the BS spectrum gives 91(99) events emplying a sizeable signal in
either case. Similar calculations can be performed in the case where the
itial photon arises from backscattered laser light.

We note in passing that a parallel process    might occur at the HERA $ep$
collider.  Firstly, if the initial $q$ or $\overline{q}$ emits a photon such
that $\gamma e$ scattering occurs we could have the reaction
$\gamma e \rightarrow E
\tilde{S}$ as above.  On the other hand, if the initial $e$ emits the
$\gamma$, then the parallel process $\gamma u \rightarrow h\tilde{E}$ or
$\gamma d \rightarrow h\tilde{S}$ can occur via the above operators in $W$.
Numerically, however, for $m_{E,h,S} \gsim 50$ GeV or so the cross sections for
these processes are quite miniscule at HERA energies ($\sqrt{s} = 314$ GeV).

Let us now turn to the reactions $gd \rightarrow \tilde{S} h$ and
$gu \rightarrow \tilde{E} h$ which can arise from the above operators.  The
subprocess differential cross section is directly obtainable from Eq.~(6) with
the replacements $\alpha^2 \rightarrow {{1} \over {6}} \alpha\alpha_s(\hat{s})$
and $m_E \rightarrow m_h$.  Placing a rapidity cut, $Y$, on the outgoing pair
implies as usual
$$
z_0 = min \left[ \beta^{-1}_h tanh (Y -|y|), 1\right]
\auto
$$
where $\beta_h \equiv \beta \left[ 1 + (m^2_h - m^2_S)/\hat{s}\right]^{-1}$
with
$\beta$ given in Eq.~(7).  Then
$$
\sigma = \int^1_{\tau_0}\, d\tau\, \int^Y_{-Y}\, dy\ \left[ g(x_a,\hat{s})
q(x_b,\hat{s}) + g(x_b, \hat{s}) q (x_a, \hat{s}) + q \rightarrow \overline{q}
\right] \int^{z_0}_{-z_0}\, {{d\sigma} \over {dz_0}}
\auto
$$
where $\tau_0 \equiv (m_h + m_S)^2/s$,  $x_{a,b} = \sqrt{\tau}\, e^{\pm y}$ and
$q$ is either $u$ or $d$; we take $Y = 2.5$ in our calculations below.  (Note
that we sum over both $gq$ and $g\overline{q}$ subprocesses in the equation
above.) $\tau$ is simply the scaled invariant mass of the h,S pair.

Figures 3a--d show the integrated cross section, $\sigma$, assuming
$\kappa=1$ as a
function of $m_h$ at both the SSC ($\sqrt{s} = 40$ TeV) and LHC
($\sqrt{s} = 15.4$
TeV) for both $gd$- and $gu$-induced processes.  As a comparison case, let us
examine the values of $\sigma$ for $m_h$ = 3 TeV and $m_S$ or $m_E = 1$ TeV.
For the $gd$
case we obtain $\sigma =
26.7 (1.21)fb$ at the SSC (LHC) and for the $gu$
case we obtain $\sigma = 5.84 (3.95) fb^{-1}$ at the SSC (LHC).  (These
numerical results were obtained using the NLO Morfin-Tung$^8$ structure
functions ,set
S1.)  Note that the $h\tilde{E}$ cross section is larger than $h\tilde{S}$ due
to the larger $gu$ luminosity in comparison to that
available for $gd$. For the SSC (LHC)
with an integrated luminosity of 10(400)fb$^{-1}$ these cross sections
translate
into a substantial event sample. If we demand that 100 events be produced at
either machine to observe the associated production of exotic particles then,
with the above luminosities,we find the following search ranges in the case
that $\kappa = 1$:

$$
\eqalign{
h\tilde{S}\left\{
\eqalign{& SSC \hspace{0.65in} m_h + m_S \lsim 4.5\ {\rm TeV}\cr
         & LHC \hspace{0.65in} m_n + m_S \lsim 3.7\ {\rm TeV}\cr}
\right.}
$$
\nskip{-.25in}
\begin{flushright}
(13)
\end{flushright}
\nskip{-.25in}
$$\eqalign{
h\tilde{E}\left\{
\eqalign{& SSC \hspace{0.65in} m_h + m_E \lsim 5.2\ {\rm TeV}\cr
         & LHC \hspace{0.65in} m_h + m_E \lsim 4.3\ {\rm TeV}\cr}
\right.
}
$$
The signature for $h,\ E$, and $S$ production themselves at supercolliders
have been discussed elsewhere.$^2$

The rates for these types of processes at the Tevatron are, of course,
substantially smaller.
For example, for $m_h = m_{S(E)} = 50$ GeV we find the total cross section to
be (now for $Y = 1.5$) $371\ pb\ (564\ pb)$ for $\kappa=1$ which is sizable for
current integrated luminosities but falls fast as both $m_h$ and
$m_{S(E)}$ increase. Clearly, however, if $\kappa$
is near unity such small masses
for the exotics must already be disfavored by the existing data.
 Figs. 4a and 4b show the cross
sections for $gd \rightarrow h\tilde{S}$ and $gu \rightarrow h\tilde{E}$ at the
Tevatron ($\sqrt{s} = 1.8$ TeV) as functions of $m_h$ for different values of
$m_{S(E)}$ and $\kappa=1$.
 Assuming as above that 100 events is necessary for a clear
discovery of exotic particles and an integrated luminosity of $100\ pb^{-1}$ we
see that the Tevatron search range for $h\tilde{E}$ and $h\tilde{S}$ final
states are approximately $m_h + m_E \lsim 330$ GeV and $m_h +m_S \lsim 290$
GeV,
respectively.  These search ranges can each be extended by $\simeq 100$ GeV if
the integrated luminosity were to reach $500\ pb^{-1}$.  Clearly, though the
search ranges are relatively restricted in comparison to the SSC/LHC, the
Tevatron can do an excellent job in the low mass range of exotics if $\kappa$
is not
too small.

In this paper we have begun to explore the collider phenomenology associated
with explicit R-parity
violating operators in $E_6$ models.  The main points of our
discussions are as follows:

\begin{itemize}
\item[{(\it i)}] We have found that the enriched particle spectrum of $E_6$
models can lead to as many as 25 trilinear and 7 bilinear operators in the
superpotential, $W$, when only $SU(3)_c \times SU(2)_L \times U(1)_Y$ symmetry
is required.  This is 16 more than is present in the standard $E_6$ model
{\it beyond}\ the usual R-parity violating terms of MSSM.  Demanding only
$|\Delta B| \neq 0$ {\it or}\ $|\Delta L| \neq 0$ terms in $W$ was found to
restrict the exotic particle quantum number assignments. Our discussion was,
however, restricted to terms which explicitly break R-parity and are
renormalizable.
\item[{(\it ii)}] Enlarging the gauge symmetry of the SM can greatly reduce
the number of new operators in $W$.  The set of allowed operators was found to
be quite sensitive to the nature of this extension. We noted that in extended
gauge models spontaneous violation of R-parity could result in the usual
R-parity breaking terms of the conventional MSSM via loop corrections even
when the extended gauge symmetries forbid them at tree level.
\item[{(\it iii)}] As in the MSSM with R-parity violation, the LSP was found
to be unstable but in our case
could only decay via virtual exotic particle exchange if
the R-parity violating operators of the MSSM were absent.  Since its lifetime
was found to scale as the eleventh power of its mass, the possibility of using
the missing energy signature for SUSY was found to be quite sensitive to the
model parameters.
\item[{(\it iv)}] The new operators explicitly
breaking R-parity in $E_6$ models allow for
new ways to produce exotic particles via gluon-quark fusion at hadron colliders
or $\gamma-e$ fusion at $e^+e^-$ colliders.  Cross sections were found to be
quite substantial over a wide range of particle masses provided the Yukawa
couplings of these new operators were not too small.
\end{itemize}

It is clear from the above that the phenomenology of $E_6$ models with R-parity
violation is particularly rich and is just beginning to be explored.

\mainhead{Acknowledgments}

The author would like to thank J. L. Hewett for discussions related to this
work.  This research has been supported by the U.S. Department of Energy,
Division of High Energy Physics, Contracts W-31-109-38 and W-7405-Eng-82.

\newpage

\mainhead{REFERENCES}
\begin{enumerate}
\item The phenomenology of R-parity violation in the MSSM is
particularly rich; see for example L.~J.~Hall and M.~Suzuki, Nuc. Phys.
{\bf B231}, 419 (1984); S.~Dawson, Nucl. Phys. {\bf B261}, 297 (1985);
S.~Dimopoulos and L.~S. Hall, Phys. Lett. {\bf B207}, 210 (1987); V.~Barger,
E.~G. Guidice, and T.~Han, Phys. Rev. {\bf D40}, 2987 (1989);
{\hbox {S.~Dimopoulos}}
\etal, Phys. Rev. {\bf D41}, 2988 (1990); H.~Dreiner and G.~Ross, Oxford
preprint OUTP-91-15P (1991); R.~Mohapatra, Phys. Rev. {\bf D34}, 3457 (1986);
R.~Barbieri and A.~Masiero, Nucl. Phys. {\bf B267}, 679 (1986), S.~Lola
and J.~McCurry, Oxford preprint OUTP-91-31P (1991); M.~C.~Gonzalez-Garcia,
J.~C.~Romao, and J.~W.~F.~Valle, Univ. of Wisconsin preprint MAD/PH/678 (1991);
E.~Ma, and D.~Ng, Phys. Rev. {\bf D41}, 1005 (1990); M.~Doncheski and
J.~Hewett, Argonne preprint (1992).
\item For a review of $E_6$ phenomenology, see J.~L.~Hewett
and T.~G.~Rizzo, Phys. Rev. {\bf 183}, 193 (1989).
\item  For a review and original references, R.~N.~Mohapatra, {\it Unification
and Supersymmetry}, (Springer, New York, 1986).
\item E. Ma, Phys. Rev. {\bf D36}, 274 (1987); Mod. Phys. Lett. {\bf A3},
319 (1988); K.~S.~Babu  \etal, Phys. Rev. {\bf D36}, 878 (1987); V.~Barger
and K.~Whisnant, Int. J. Mod. Phys. {\bf A3}, 879 (1988); J.~F.~Gunion \etal,
Int. J. Mod. Phys. {\bf A2}, 118 (1987); T.~G.~Rizzo, Phys. Lett. {\bf
B206}, 133 (1988).
\item C. Hattori \etal, Nagoya University preprints DPNU-91-21 (1991) and
DPNU-91-32 (1991).
\item For a review of the EPA and original references, see P.~W.~Johnson,
F.~Olness, and W.-K.~Tung in {\it Proceedings of the 1986 Summer Study on the
Physics of the Superconducting Supercollider}, Edited by R.~Donaldson and
J.~Marx (1986).
\item R. Blankenbecler and S. Drell, Phys. Rev. Lett. {\bf G1}, 2324 (1988),
Phys. Rev. {\bf D36}, 277 (1987) and {\bf D37}, 3308 (1988),
D.~V.~Schroeder, SLAC-Report-371 (1990); F.~Halzen, C.~S.~Kim and M.~Stong,
Univ. Of Wisconsin preprint MAD/PH/673 (1991).
\item J. C. Morfin and W.-K. Tung, Z. Phys. {\bf C52}, 13 (1992).
\end{enumerate}

\newpage

\mainhead{TABLE CAPTIONS}
\begin{itemize}
\item[{Table 1.}] Nomenclature and $SU(3)_c \times SU(2)_L \times U(1)_Y$
quantum numbers for the fields in the \underline {27}\ representation of $E_6$.
\item[{Table 2.}] The 25 trilinear terms allowed in the $E_6$ superpotential
$W$ by $SU(3)_c \times SU(2)_L \times U(1)_Y$ gauge invariance following the
notation of Table~1.  HR labels the term as it appears in $W$
from the review of
Hewett and Rizzo (ref.~2). $\theta$ labels the value of the mixing
angle between $Z_\psi$ and $Z_\chi$ described in the text which allows this
particular term: $\chi(\theta = -90^\circ)$, $\eta(\theta \simeq 37.76^\circ)$,
$\nu_R(\theta \simeq -14.48^\circ)$, $\nu^I_R(\theta \simeq -66.72)$,
$I(\theta \simeq -52.24^\circ)$, and $-\eta(\theta = -37.76^\circ)$.  We
also show which terms generate particle masses, induce leptoquark ($L$) or
diquark ($D$) quantum numbers for $h$ and which terms appear in the R-parity
violating version of the MSSM.  Also shown is whether the term is allowed in
either the LRM or ALRM case.
\item[{Table 3.}] Same as Table 2 but for the bilinear terms in $W$ allowed
by $SU(3)_c \times SU(2)_L \times U(1)_Y$ gauge invariance.  Also shown is
whether these terms are allowed in the LRM or ALRM.
\end{itemize}

\newpage

\centerline{Table 1}
\begin{center}
\begin{tabular}{|c|c|c|c|c|c|c|} \hline
SO(10) & SU(5) & { } & Color & $T_{3L}$ & Y/2 & Q \\
{} & {} & {} & {} & {} & {} & {} \\ \hline \hline
16 & 10 & $Q\equiv \left( {{u} \atop {d}} \right)_L$ & 3 &
$\left( {{\phantom{-}1/2} \atop {-1/2}} \right)$ & 1/6 &
$\left( {{2/3}\atop {1/3}}\right)$ \\
{} & {} & {} & {} & {} & {} & {} \\ \hline
{} & {} & $u^c_L$ & $\overline{3}$ & 0 & -2/3 & -2/3 \\
{} & {} & {} & {} & {} & {} & {} \\ \hline
{} & {} & $e^c_L$ & 1 & 0 & 1 & 1 \\
{} & {} & {} & {} & {} & {} & {} \\ \hline
{} & $\overline{5}$ & $L\equiv \left( {{\nu} \atop {e}}\right)_L$ & 1 &
$\left( {{\phantom{-}1/2} \atop {-1/2}}\right)$ & -1.2 &
$\left( {{\phantom{-}0} \atop {-1}}\right)$ \\
{} & {} & {} & {} & {} & {} & {} \\ \hline
{} & {} & $d^c_L$ & $\overline{3}$ & 0 & 1/3 & 1/3 \\
{} & {} & {} & {} & {} & {} & {} \\ \hline
{} & 1 & $\nu^c_L$ & 1 & 0 & 0 & 0 \\
{} & {} & {} & {} & {} & {} & {} \\ \hline
10 & $\overline{5}$ & $H \equiv \left( {{N} \atop {E}}\right)_L$ &
1 & $\left( {{\phantom{-}1/2} \atop {-1/2}}\right)$ & -1/2 &
$\left( {{\phantom{-}0} \atop {-1}}\right)$ \\
{} & {} & {} & {} & {} & {} & {} \\ \hline
{} & {} & $h^c_L$ & $\overline{3}$ & 0 & 1/3 & 1/3 \\
{} & {} & {} & {} & {} & {} & {} \\ \hline
{} & 5 & $H^c\equiv \left( {{E} \atop {N}}\right)^c_L$ & 1 &
$\left( {{\phantom{-}1/2} \atop {-1/2}}\right)$ &
1/2 & $\left( {{1} \atop {0}}\right)$ \\
{} & {} & {} & {} & {} & {} & {} \\ \hline
{} & {} & $h_L$ & 3 & 0 & -1/3 & -1/3 \\
{} & {} & {} & {} & {} & {} & {} \\ \hline
1 & 1 & $S^c_L$ & 1 & 0 & 0 & 0 \\
{} & {} & {} & {} & {} & {} & {} \\ \hline
\end{tabular}
\end{center}

\textheight 9.0in
\newpage
\centerline{Table 2}
\begin{center}
\begin{tabular}{@{\extracolsep{0.25in}}|r|c|r|c|c|} \hline
{} & Term & HR & $\theta$ & Comment \\ \hline \hline
1 & $QQh$ & 9 & {} & D \\ \hline
2 & $QLD^c$ & {} & $\eta$ & ${\rm MSSM}\ \st{R}\ |\Delta L| = 1$ \\ \hline
3 & $QLh^c$ & 7 & {} & L \\ \hline
4 & $Qh^cH^c$ & 1 & {} & $u\ mass$ \\ \hline
5 & $Qd^cH$  & 2 & {} & $d\ mass$ \\ \hline
6 & $Qh^cH$ & {} & $\eta$ & \\ \hline
7 & $LLe^c$ & {} & $\eta$ & ${\rm MSSM}\ \st{R}\ |\Delta L| = 1$ \\ \hline
8 & $LHe^c$ & 3 & {} & $e\,mass$ \\ \hline
9 & $LH^c\nu^c$ & 11 & {} & $Dirac\ \nu\ mass$ \\ \hline
10 & $LH^cS^c$ & {} & $\eta$ & {} \\ \hline
11 & $u^cd^cd^c$ & {} & $\eta$ & ${\rm MSSM}\ \st{R}\ |\Delta B| = 1$ \\ \hline
12 & $u^cd^ch^c$ & 10 & {} & $D$ \\ \hline
13 & $u^ce^ch$ & 6 & {} & $L$ \\ \hline
14 & $u^ch^ch^c$ & {} & $\eta$ & {} \\ \hline
15 & $d^c\nu^ch$ & 8 & {} & $L$ \\ \hline
16 & $d^cS^ch^c$ & {} & $\eta$ & {} \\ \hline
17 & $e^cHH$ & {} & $\eta$ & {} \\ \hline
18 & $(\nu^c)^3$ & {} & $\nu_R$ & $ALRM$ \\ \hline
19 & $\nu^c\nu^cS^c$ & {} & $-\eta$ & {} \\ \hline
20 & $HH^c\nu^c$ & {} & $\eta$ & {} \\ \hline
21 & $hh^c\nu^c$ & {} & $\eta$ & {} \\ \hline
22 & $\nu^cS^cS^c$ & {} & $\nu^I_R$ & {} \\ \hline
23 & $HH^cS^c$ & 4 & {} & $H\ mass$ \\ \hline
24 & $hh^cS^c$ & 5 & {} & $h\ mass$ \\ \hline
25 & $(S^c)^3$ & {} & $\chi$ & $LRM$ \\ \hline
\end{tabular}
\end{center}

\textheight 8.0in
\newpage
\centerline{Table 3}
\begin{center}
\begin{tabular}{@{\extracolsep{0.25in}}|c|c|c|c|c|} \hline
{} & Term & $\theta$ & Comment & {} \\ \hline \hline
1 & $LH^c$ & $\nu_R$ & ${\rm MSSM}\ |\Delta L| = 1$ & $ALRM$ \\ \hline
2 & $hd^c$ & $\nu_R$ & {} & $ALRM$ \\ \hline
3 & $(\nu^c)^2$ & $\nu_R$ & {} & $ALRM$ \\ \hline
4 & $\nu^cS^c$ & I & {} & {} \\ \hline
5 & $HH^c$  & $\chi$ & MSSM & $LRM$ \\ \hline
6 & $hh^c$ & $\chi$ & {} & $LRM$ \\ \hline
7 & $(S^c)^2$ & $\chi$ & {} & $LRM$ \\ \hline
\end{tabular}
\end{center}

\newpage

\mainhead{FIGURE CAPTIONS}

\begin{itemize}
\item[{Fig. 1.}] Decay length in $cm$ for the $LSP(\chi)$
as a function of $m_\chi$
assuming $\Lambda = 100$ GeV and that $\chi$ is produced with an energy of 100
GeV.  The bottom curve corresponds to $\lambda =1$ with subsequently
higher curves
corresponding to a decrease in $\lambda$ by a power of 10.
\item[{Fig. 2.}] Cross section for $\tilde{S}E^\pm$ production in high energy
$\epm$\ collisions as a function of $m_E$ with $\kappa=1$ and different $m_S$
choices.  (a) $\sqrt{s} = 500$ GeV using the EPA, (b) $\sqrt{s} = 500$ GeV
using BS.  The top curve in both cases corresponds to $m_S = 50$ GeV and each
subsequently lower curve corresponds to an increase in $m_S$ by 50 GeV.  (c)
and (d) same as (a) and (b) but for a $\sqrt{s} = 1$ TeV collider.  The top
curve corresponds to $m_S = 100$ GeV and each subsequently lower curve
corresponds to an increase in $m_S$ by 100 GeV.
\item[{Fig. 3.}] Cross section after rapidity cuts for $h\tilde{S}$
[(a) and (b)] or $h\tilde{E}$ [(c) and (d)] production at hadron supercolliders
for $\kappa=1$
as a function of $m_h$ assuming different values for $m_S$ or $m_E$.
(a) $h\tilde{S}$ production at the SSC; (b) $h\tilde{S}$ at the LHC, (c)
$h\tilde{E}$ production at the SSC; (d) $h\tilde{E}$ production at the LHC.
The top curve corresponds to $m_{S(E)} = 0.1$ TeV with each successive
curve corresponding to an increase in $m_{S(E)}$ by 0.1 TeV.
\item[{Fig. 4.}] The $h\tilde{E}$ (a) and $h\tilde{S}$ (b) production
cross section
as a function of $m_h$ with $\kappa=1$ at the Tevatron.  The top solid curve
corresponds to $m_{E,S} = 50$ GeV with lower subsequent curves corresponding to
$m_{E,S} = 100,\ 200,\ 300,$ etc. GeV.
\end{itemize}

\end{document}